\documentstyle[12pt]{article}
\newfont{\bmits}{cmmib10}  % this is for subscript - replace as appropriate
\newcommand{\bmit}[1]{\mbox{\boldmath $#1$}}

\begin{document}

\baselineskip24pt

\begin{center}
{\large \bf Chaos induced by Pauli blocking}

\vskip 1truecm

S.~Dro\.zd\.z$^1$, J.~Oko\l{}owicz$^1$, M.~Ploszajczak$^2$,
E.~Caurier$^3$ and T.~Srokowski$^{2}$
\vskip 1truecm

$^1$ {\em Institute of Nuclear Physics, PL -- 31-342 Krak\'ow, Poland}

$^2$ {\em  GANIL, BP~5027, F-14021 Caen Cedex, France}

$^3$ {\em Centre de Recherches Nucl\'eaires, F-67037 Strasbourg Cedex, France}

\vskip 2truecm

{\large ABSTRACT}

\end{center}

Dynamics of classical scattering in the system of fermions is studied.
The model is based on the coherent state representation
and the equations of motion for fermions
are derived from the time-dependent
variational principle.
It is found that the antisymmetrization due to the Pauli exclusion
principle, may lead to hyperbolic chaotic scattering even in the absence
of interaction between particles. At low bombarding energies,
the same effect leads to the screening of the hard,
short-ranged component in the two particle interaction and thus
regularizes the dynamics.

\vskip 1truecm
\noindent {\bf PACS:} 05.45.+b; 05.60.+w; 24.10.-i

\vfill
\newpage

One of the central issues in the theory of complex dynamical systems is
the classical-quantum correspondence for classically chaotic motion.
Classically, chaos is a well defined concept but a way it manifests
itself on the quantum level still remains a matter of controversy.
One of the reasons of this controversy is the
quantum mechanical symmetry related to the identity of particles.
Of particular interest and importance in this context are
the effects resulting from antisymmetry
of an underlying wave function for fermionic systems.
Most of the physical objects, like atoms
or atomic nuclei,
providing empirical input for the above studies, are collections of
fermions. The fundamental question
then is in which sense does a fermionic nature of
the basic constituents influence
the structure of the corresponding classical
phase space. And, in particular, does it increase or reduce the amount of
instabilities leading to chaotic behavior?

The model which seems best
suited to incorporate the elements needed and thus
for addressing such questions
is the one based on the coherent state formalism.
This formalism allows to define a classical limit~\cite{Yaf} and,
at the same time, to accommodate the effects of antisymmetrization.
Consistently, the $i$-th single particle
state is described by the Gaussian wave packet~:
\begin{equation}
\phi_i({\bmit r}) \equiv
\langle {\bmit r}\vert \phi_{{\mbox{\bmits Z}}_i}\rangle =
\left( {1 \over \pi b^{2}}\right)^{3/4}
\exp \left[-\frac{({\bmit r}-{\bmit Z}_i)^2}{2 b^{2}}\right]
\label{eq:sp}
\end{equation}
where ${\bmit Z}_i$ is the location of the center of gravity of the packet.
These single particle states are not mutually orthogonal and, thus,
their overlaps
$n_{ij}=\langle \phi_i \vert \phi_j \rangle =
\exp \left[- ({{\bmit Z}_i}^{*} - {\bmit Z}_j)^{2}/(4b^{2})\right]$
do not vanish.
The $A$-fermion system is then described by a Slater determinant
$\Phi ({\bmit Z}) = (A!)^{-1/2} \det [\phi_i({\bmit r}_j)]$
whose norm
$N({\bmit Z})$ is equal to $\det [ n_{ij} ]$.
In this way the state of the system
is entirely specified by the parameters
${\bmit Z} \equiv \{ {\bmit Z}_i: {i=1,\ldots,A} \}$
which for a proper treatment of the dynamics have to be considered
as complex $({\bmit Z}={\bmit R}+i{\bmit P})$~\cite{KK}.
The units are specified by setting
$\hbar=b=m=\hbar \omega=1$.
The time development of the dynamical variables can be determined by the
time-dependent variational principle:
\begin{equation}
\delta \int_{t_1}^{t_2} dt \langle \Phi({\bmit Z}) \vert (i {d\over dt} -
{\hat H}) \vert \Phi({\bmit Z}) \rangle N({\bmit Z})^{-1} =0.
\end{equation}
The resulting equations of motion take the form:
\begin{equation}
i \sum_{j\beta}^{}
 S_{i\alpha,j\beta}{\dot Z}_{j\beta} =
{\partial H \over \partial Z^*_{i\alpha}}
\label{eq:em}
\end{equation}
where $\alpha, \beta = x, y, z$.
$H$ is the expectation value of the many body Hamiltonian
$H({\bmit Z}^*,{\bmit Z}) = \langle \Phi({\bmit Z})\vert {\hat H}\vert
\Phi({\bmit Z}) \rangle N({\bmit Z}^*,{\bmit Z})^{-1}$
and the Hermitian matrix:
\begin{equation}
S_{i\alpha,j\beta} =
{\partial^2 \over \partial Z^*_{i\alpha} \partial Z_{j\beta}} \log N .
\end{equation}
is positive definite.
Such a scheme constitutes a formal basis for various molecular dynamics
approaches~\cite{DOP,AS,OHM}
and the Gaussian form of the wave packet proves
appropriate for semiclassical studies of the structure of the phase
space~\cite{Hel}.

In general, due to the antisymmetrization,
the off diagonal terms in $S_{i\alpha, j\beta}$ do not disappear.
Therefore, neither
${\bmit Z}_i$ and ${\bmit Z}^*_i$ nor their
real ${\bmit R}_i$ and imaginary
${\bmit P}_i$ parts form the canonically conjugate variables.
For the two particle system,
an exact transformation can be performed to the
new, canonically conjugate variables ${\bmit W}_i~{(i = 1,2)}$~\cite{SKF}.
These ${\bmit W}_i$ cannot get closer
than $\sqrt{2}$ independently of the difference
between ${\bmit Z}_1$ and ${\bmit Z}_2$.
Thus a topological hole, existing also for $A \ge 3$,
corresponds to the Pauli forbidden region~\cite{SKF,horil}.
However, an explicit expression for the canonical variables
is not known for $A\ge 3$ and, therefore, the
equations~(\ref{eq:em}) will be solved in original variables ${\bmit Z}$.

Schematic models~\cite{JS,GR,DOS},
based on the scattering of a particle
on the target composed of three particles
at the corners of equilateral triangle located in the reaction
plane, proved very instructive in studying various
aspects of the collision processes.
In this case, the scattering problem reduces
to two degrees of freedom.
At the present exploratory stage we perform
an analogous study.
The target is given by a three
fermion (Gaussian wave packets (\ref{eq:sp}))
configuration forming equilateral
triangle of side equal to $4$.
We begin by entirely discarding the interaction term
in the Hamiltonian
in order to elucidate on the role of antisymmetrization itself.
The Pauli forbidden regions manifest themselves by the strong increase of
$H({\bmit Z}^*,{\bmit Z})$ generated by the kinetic energy operator.
Equivalently, the form of the energy
surfaces in $\bmit R$ depends on the momentum
like variables $\bmit P$, as demonstrated in Fig.~1.
Here the $\bmit R$ dependence of the
energy, as `seen' by the fourth fermion
located in the reaction plane $(x,y)$, is plotted
at two fixed values of $\bmit P$.
With increasing $\bmit P$, the hilly structures in this
effective Pauli potential
become relatively smaller and the dynamics approach
a free motion. Notice that a $\bmit P$ dependence of the
Pauli potential is an analog of the nonlocality
for ordinary potentials.

In a real dynamical process $\bmit P$ changes in time
and so does $H$. Nevertheless, the appearance of the three center
structures suggests~\cite{JS,GR,DOS} that for certain $\bmit P$ values one
may expect a strong
sensitivity on the initial conditions.
That this really happens is
documented in Fig.~2 which shows, for the three different
energies, the impact parameter dependence of the
deflection angle $\theta$ of the particle scattered off the
target.  The motion is initialized
by setting the appropriate initial
value of $P_x$ with $P_y=0$ and the impact parameter $R_y$.
The presentation of the results is based
on the asymptotic values of the variables in the region where
antisymmetrization is no longer effective
and, thus, $\bmit R$, $\bmit P$  are canonical conjugate.
For energies either small $(E=0.2)$ or large $(E=0.6)$, compared
to the height of the effective Pauli potential
$V_{\mbox{\scriptsize max}}({\bmit P}=0) \equiv \max H({\bmit R},{\bmit P}=0)
\simeq 1/2$ (see Fig.~1a),
$\theta$~~depends on the impact parameter but this dependence
is essentially continuous.
At $E=0.4$  and for $R_y$ between 0 and 1, one
observes behavior characteristic of the chaotic scattering.

Very interestingly, this scattering process carries all characteristics
of the hyperbolic chaotic scattering~\cite{ER}.
For this type of scattering, theory predicts~\cite{ER,KG,DOS}
that the survival probability, {\em i.e.}\ number
$N(t)$ of trajectories remaining in
the interaction region up to time $t$, follows:
\begin{equation}
N(t)=N_0 \exp [-\lambda \,(1-D)\, t]\, ,
\label{eq:sur}
\end{equation}
where $\lambda$ is the Lyapunov exponent and $D$ is the fractal
dimension. The Lyapunov exponent
$\lambda$, calculated from the growth rate of separation ratio
$\delta/\delta_0$ between the two neighboring
(and long enough) trajectories~\cite{ben},
is independent on what pair of scattering
trajectories is used and, thus, $\lambda = 0.22$ is well determined (Fig.~3a).
Furthermore, the set of
singularities seen in Fig.~2 for $E=0.4$ possesses a well defined fractal
dimension $D$. This one can conclude from calculation based
on the uncertainty exponent technique~\cite{HOG} which gives $D=0.591$
(Fig.~3b). Finally, following the concepts of the transport theory~\cite{Mei},
by uniform random sampling of the whole interval
of impact parameters one determines the survival probability $N(t)$.
In our case one
observes (Fig.~3c) asymptotically exact exponential dependence,
characteristic of the hyperbolic chaotic scattering,
perfectly described by the above values of $\lambda$
and $D$.

These investigations provide convincing
evidence that the chaotic scattering
we deal with is of the hyperbolic type
where the set of singularities is connected to the existence
of only unstable periodic orbits.
In the present case, however,
these structures take place for a system of {\it noninteracting fermions}
and are entirely due to the antisymmetrization.
Independently, it is interesting to notice the appearance of such uniform
fractal set of singularities for a nonlocal problem.

The most important result of the above analysis is that
for a sufficiently dense system, the correlations
resulting from Pauli blocking may lead
to a significant modification of the particle dynamics.
In certain situations, they may even convert free motion
in the gas of non-interacting particles into
the strongly chaotic one.
In general, this is more likely to occur when the mean kinetic
energy of particles
in the gas is comparable with the height of the
effective Pauli potential.
At very low energies, the range of Pauli blocking
for sufficiently dense system extends to such
a size that different topological holes
start overlapping and the dynamics becomes regular again
(see upper panel of Fig.~2). In more realistic case of
{\it interacting fermions}, this effect may thus screen
out the short range components of the two-body interaction.
This dynamical screening may be
especially important for the hard, short-range
interactions like the one between the nucleons. Classically,
such an interaction generates strongly chaotic behavior. However,
its appearance inside the Pauli forbidden region may restructure and even
eliminate the corresponding irregularities.

For a somewhat more quantitative illustration of this point,
we present in Fig.~4
the energy surface plots in the similar configuration as in Fig.~1.
This time, however,
the constituents interact with the spherically symmetric,
repulsive two-body interaction $V(r)=V_0 \exp [-(r/r_0)^2]$,
with $V_0=10^{5/2}$ and $r_0=10^{-1/2}$. For these parameters
the interaction is comparatively
hard and short-ranged. Upper panel of Fig.~4
corresponds to the situation with no antisymmetrization
included. In (a) the side
of the triangle equals $3$ and in (d) it equals $4$.
In both cases the three hill structure shows up and, consequently,
the scattering will be chaotic in the corresponding energy intervals.
Including antisymmetrization
changes the picture completely (see Fig.~4b).
Here, not only the energy is reduced by almost
two orders of magnitude but also the three hill structure, previously
responsible for chaotic behavior, disappears. Increasing momentum, slowly
recovers the original shape of the
energy surfaces, as is shown in Fig.~4 (c)
and (f) for $P_x=2\sqrt{10}$, but they
still remain about one order of magnitude lower.

Returning to our introductory question we
thus conclude that the fermionic
nature of particles may drastically change
the structure of the corresponding
classical phase space. As the two extreme
possibilities we identify the chaotic
behavior in absence of any interaction and
the regularization of
motion for strongly interacting particles.
The first of them is very intriguing in view, for instance, of the fact
that the hyperbolic chaotic scattering is
considered as a classical
manifestation of Ericson fluctuations~\cite{BS}.
Equally interesting from the physical point of view is the second effect.
The dynamical screening of the short-range components of
the two-body interaction,
is consistent with the success of the mean field
concept in the region close to the Fermi surface for such a strongly
interacting system as an atomic nucleus.
In this connection, the space
nonlocalities are identified~\cite{NY} as a crucial element enhancing the
nucleon mean free path.
Many related questions remain a subject for further
investigations.

This work was partly supported by KBN Grant No.\ 2~P302~157~04.
Two of us (S.D. and J.O.)
wish to thank GANIL for the hospitality extended to them, when
the work has been completed.

\newpage
\vspace{.25in}
\parindent=.0cm               %align refs

\newpage
\begin{center}
{\bf Figure captions}
\end{center}

\begin{itemize}
\item[{\bf Fig.~1}]
Energy surface plots for noninteracting configuration generated
by the three fermions located at the corners of equilateral triangle
as seen by the fourth
fermion at momentum $P_x=0$ (a) and $P_x=\sqrt{10} $ (b).
Triangle side length equals $4$ and $P_y=0$.

\item[{\bf Fig.~2}]
Deflection angle $\theta$ as a function of the impact parameter for three
different energies under geometrical conditions corresponding to Fig.~1.

\item[{\bf Fig.~3}]
(a) Separation ratio
between the neighboring trajectories for the scattering
process at $E=0.4$. Dots denote the dynamically determined values and the
straight solid line represents a fit whose slope corresponds to the
Lyapunov exponent $\lambda=0.22$.\par
(b) Dependence of the fraction $f(\epsilon)$ of uncertain pairs of
trajectories (differing in $\theta$ by more than $\pi /2$)
as a function of the difference $\epsilon$ in initial values
of the impact parameters. According to the
uncertainty exponent technique
the corresponding fractal dimension is $D=0.591$.
\par
(c) Survival probability expressed
as a number of the scattering trajectories
remaining in the interaction region up to time $t$. The straight line
represents the theoretically determined dependence.

\item[{\bf Fig.~4}]
Energy surface plots for the three
fermion configurations forming equilateral
triangle of side equal to 3 (a, b, c)
and 4 (d, e, f) respectively: (a) and (d) represent
the potential energy without antisymmetrization
for the two-body interaction defined in the text,
(b) and (e) the total energy with antisymmetrization,
(c) and (f) the total
energy with $P=2\sqrt{10}$

\end{itemize}

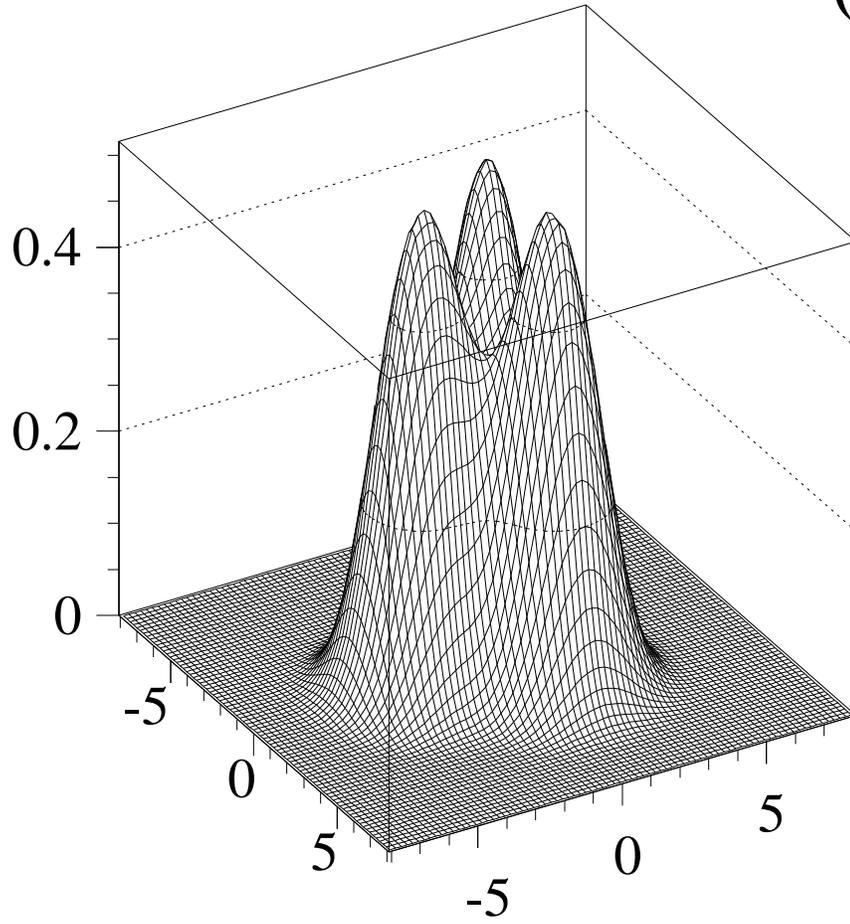
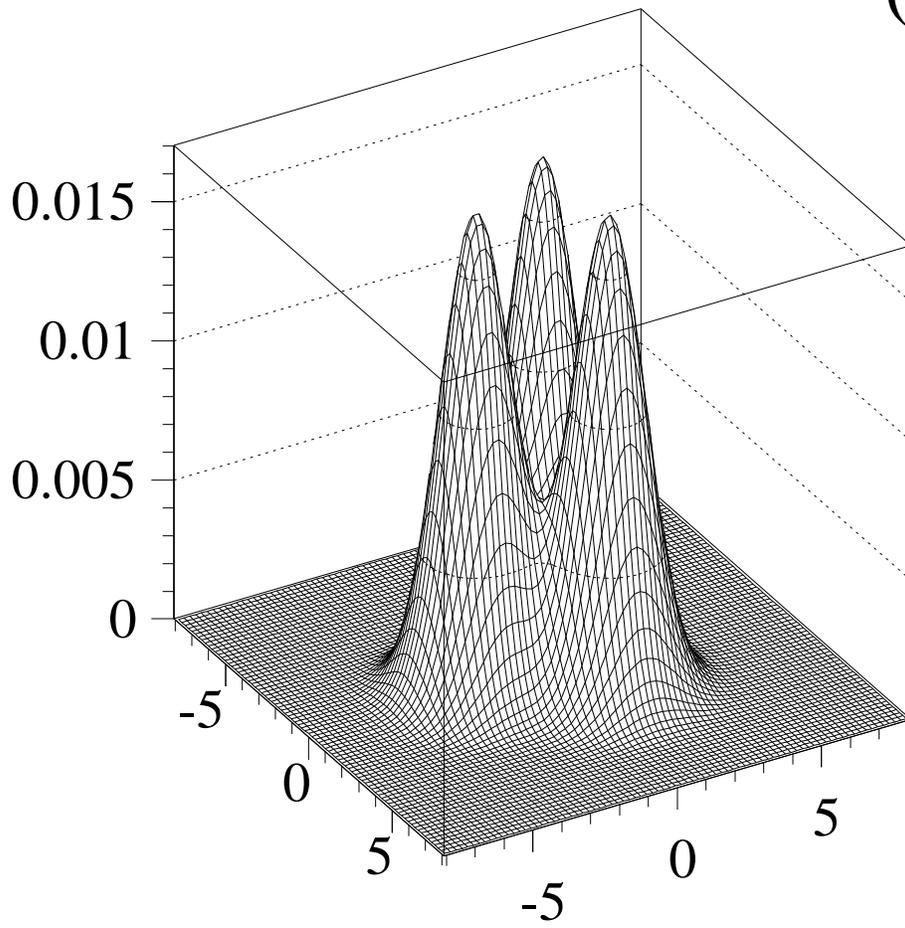
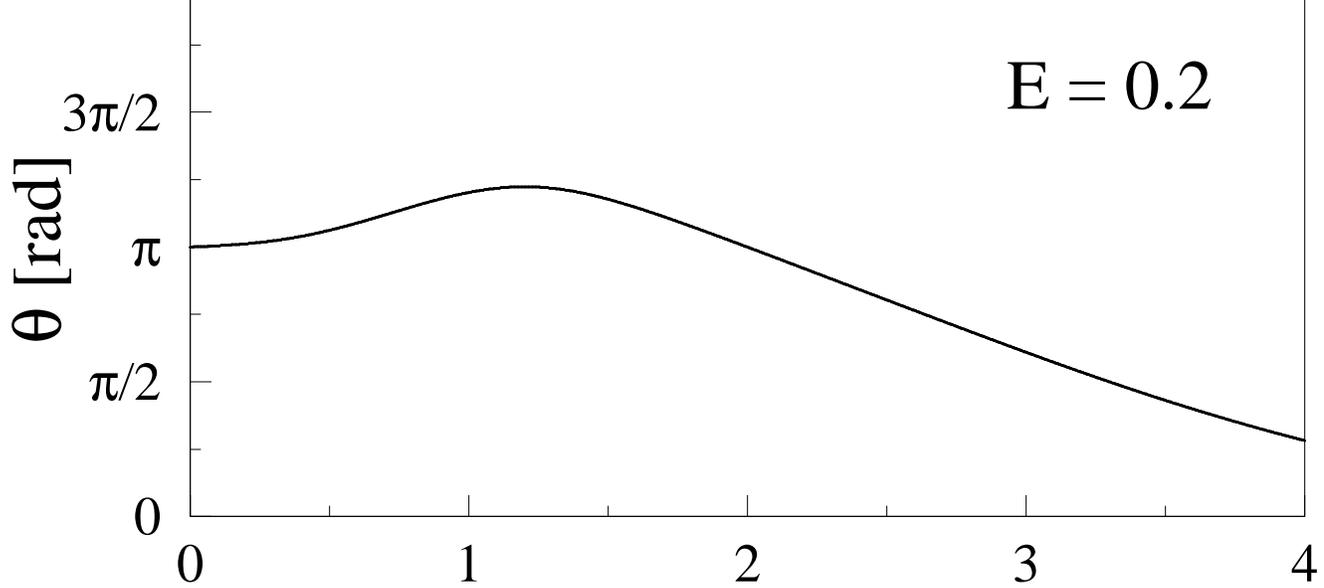
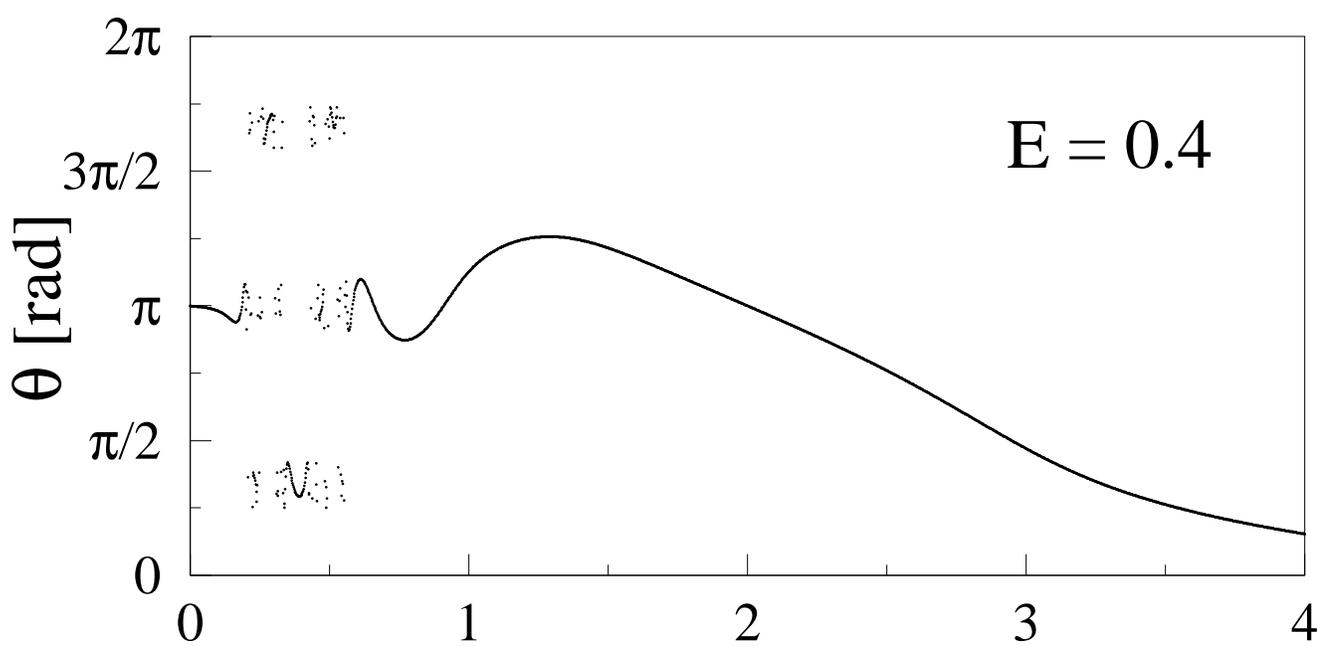
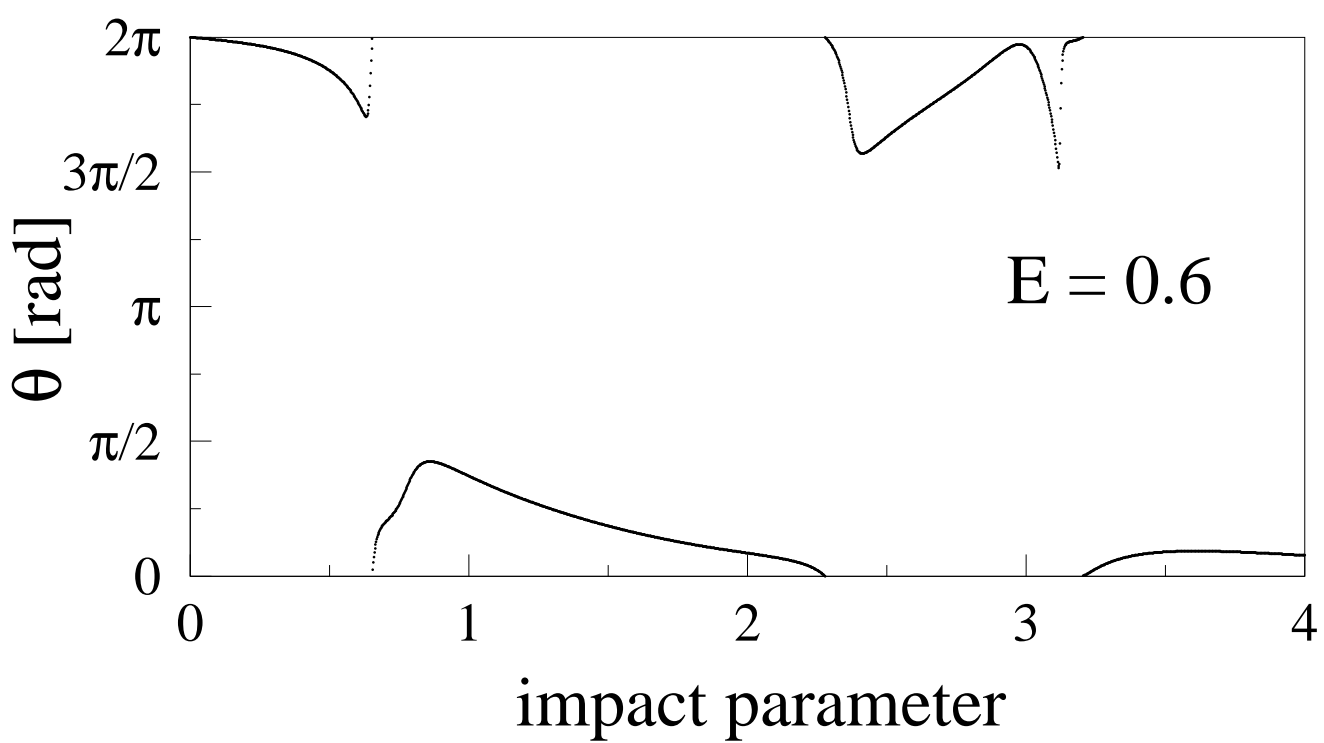
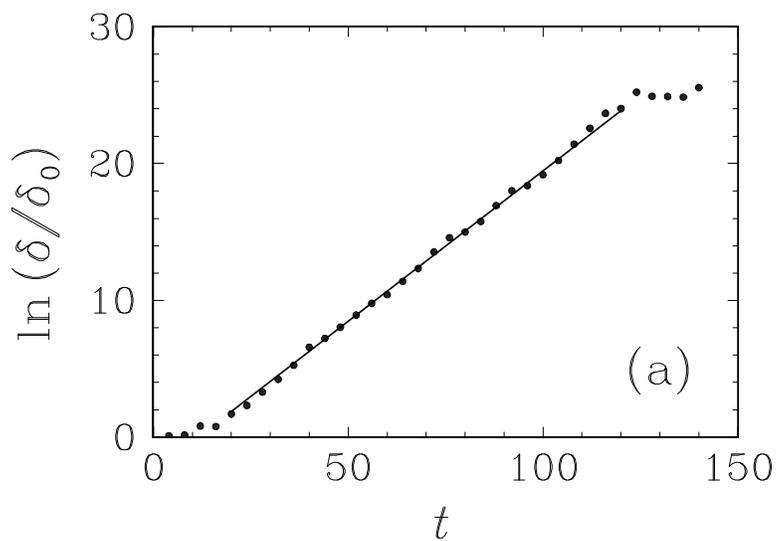
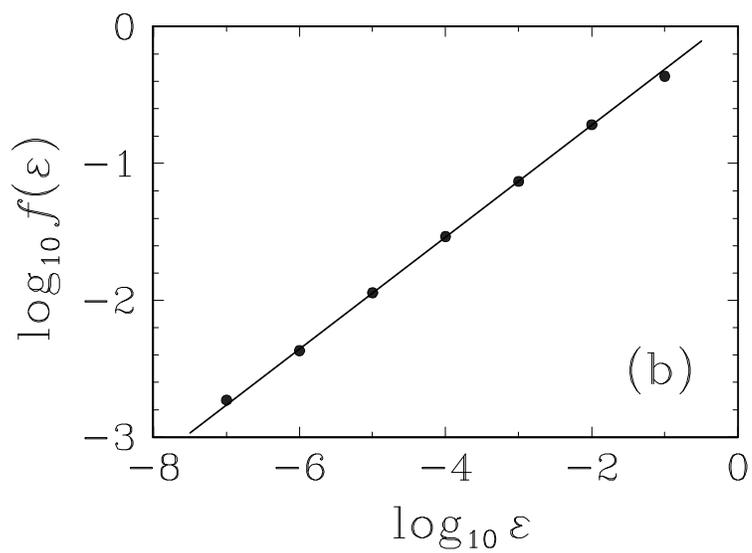
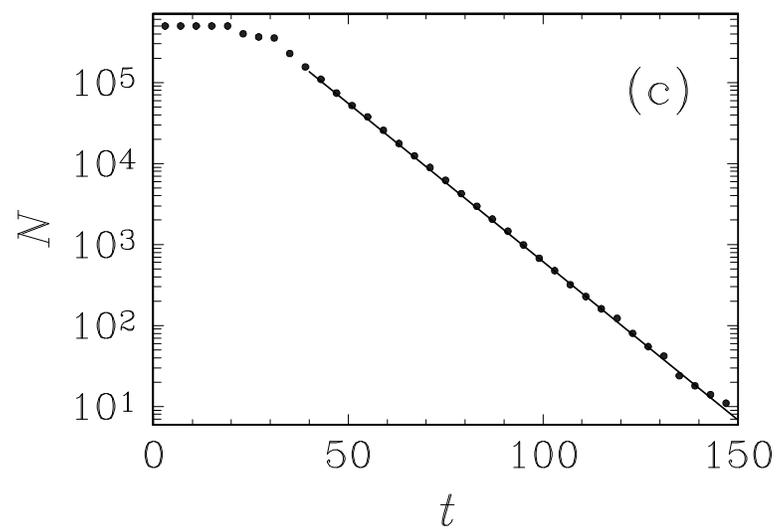
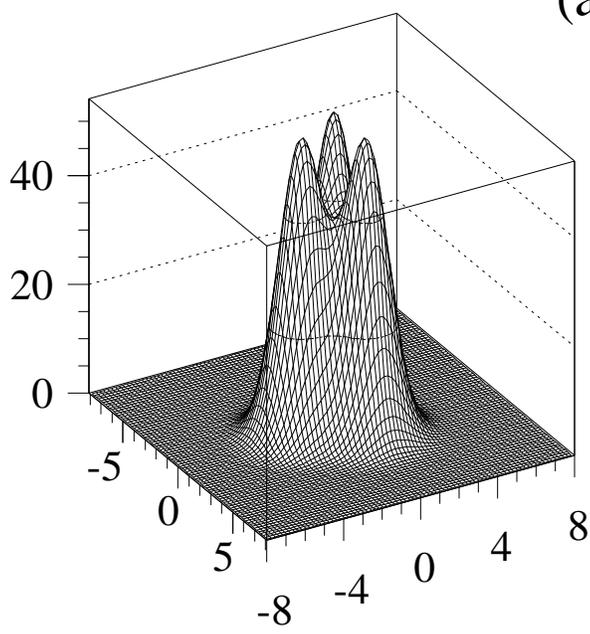
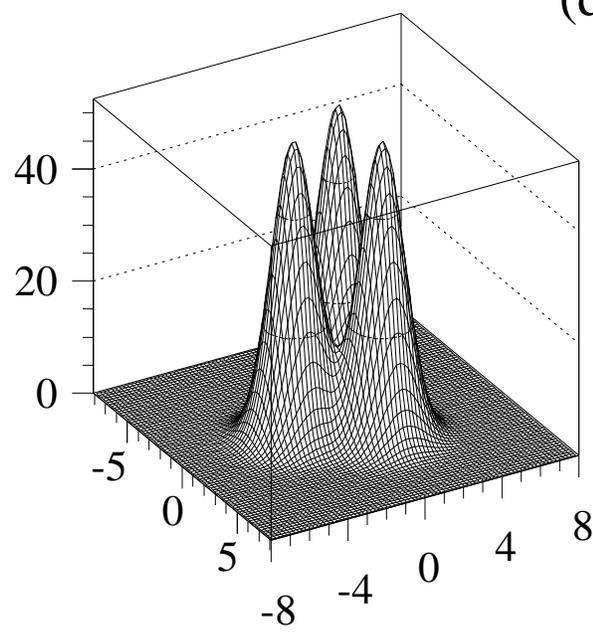
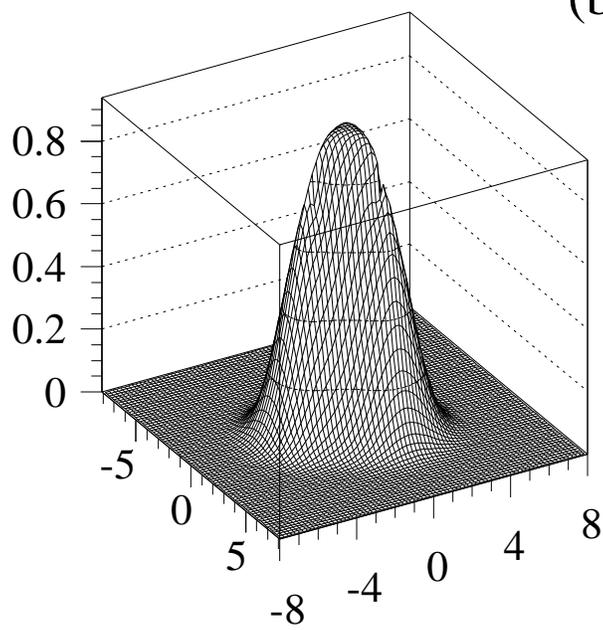
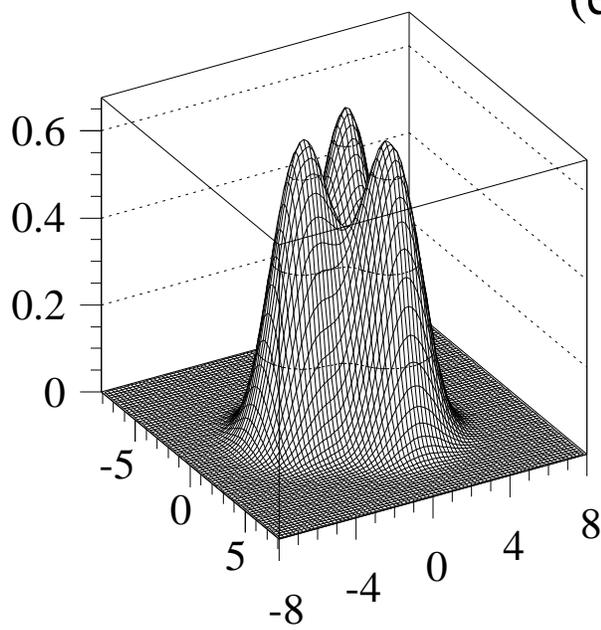
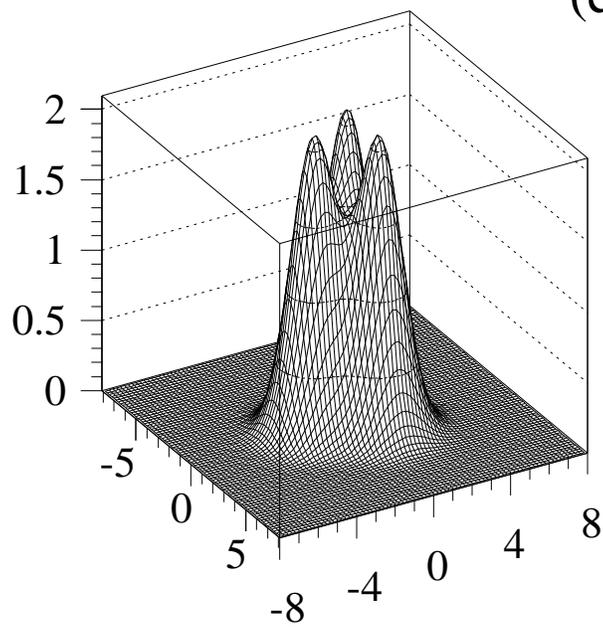
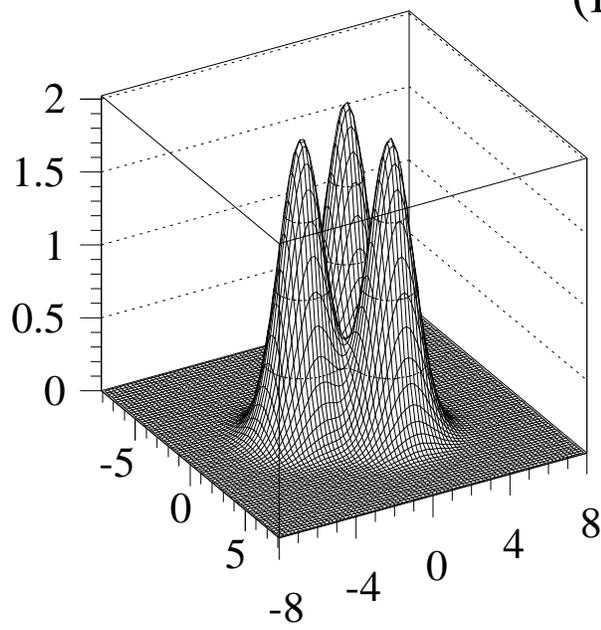
\end{document}